\documentclass{jfm}
\usepackage{graphicx}
\usepackage{epstopdf, epsfig}
\usepackage{subfig,setspace}
\usepackage{color}
\usepackage{rotating}
\usepackage{pdflscape}
\usepackage{amsmath}
\usepackage{lscape}

\shorttitle{Impact of domain anisotropy}
\shortauthor{K. Julien et al}

\title{Impact of domain anisotropy on the inverse cascade in geostrophic turbulent convection}

\author{
 Keith Julien\aff{1} \corresp{\email{julien@colorado.edu}},
 Edgar Knobloch\aff{2}
  \and 
Meredith Plumley\aff{1}
}

\affiliation{\aff{1}Department of Applied Mathematics, University of Colorado, Boulder, CO 80309, USA
\aff{2}Department of Physics, University of California, Berkeley, CA 94720, USA }

\begin{document}

\maketitle

\begin{abstract}

The effect of domain anisotropy on the inverse cascade occurring within the geostrophic turbulence regime of rapidly rotating Rayleigh-B\'enard convection (RRBC) is investigated. In periodic domains with square cross-section in the horizontal a domain-filling dipole state is present. For rectangular periodic domains a Kolmogorov-like flow consisting of a periodic array of alternating unidirectional jets with embedded vortices is observed, together with an underlying weak meandering transverse jet. Similar transitions occurring in weakly dissipative two-dimensional flows driven by externally imposed small amplitude noise as well as in classical hydrostatic geostrophic turbulence are a consequence of inviscid conservation of energy and potential enstrophy and can be understood using statistical mechanics considerations. {RRBC represents an important three-dimensional system with only one inviscid invariant that nonetheless exhibits large-scale structures driven by intrinsically generated fluctuations.}  
\end{abstract}

\begin{keywords}
..
\vspace{-3ex}
\end{keywords}

\section{Introduction}\label{sec:intro}
The quintessential paradigm for investigating the fundamentals of rotating, thermally forced flows is provided by rotating Rayleigh-B\'enard convection (RRBC) in a horizontal layer rotating about a vertical axis with constant angular velocity $\Omega$, i.e., convection in a layer of Boussinesq fluid confined between flat, horizontal, rigidly rotating upper and lower boundaries maintaining a destabilizing temperature jump $\Delta T>0$. Of particular relevance to the dynamics of stellar and planetary interiors, planetary atmospheres and terrestrial oceans is the regime of geostrophic turbulence where fluid motions are sufficiently constrained by rotation to enforce pointwise balance between the pressure gradient and the Coriolis force, otherwise known as geostrophic balance. This balance is characteristic of rapidly rotating systems for which the convective Rossby number is small:
\begin{equation}
\label{eqn:nond0}
Ro\equiv \sqrt{\frac{g\alpha \Delta T H}{2\Omega}} = \sqrt{\frac{Ra}{Pr}} E \ll 1 \,.
\end{equation}
This number denotes the ratio of the rotation timescale to the free-fall or free-rise timescale for a parcel of fluid with temperature difference $\Delta T$ relative to the ambient fluid. Here $g$ denotes acceleration due to gravity, $\alpha$ is the thermal expansion coefficient and $H$ is the layer depth. The second equality rewrites this definition in terms of quantities familiar from studies of Rayleigh-B\'enard convection: the Rayleigh number $Ra$, Ekman number $E$ and the Prandtl number $Pr$ given by
\begin{equation}
\label{eqn:nondb}
Ra=\frac{g\alpha \Delta T H^3}{\nu \kappa},\qquad 
E=\frac{\nu}{2 \Omega H^2},\qquad
Pr=\frac{\nu}{\kappa} \, .
\end{equation}
These measure, respectively, the strength of the thermal forcing, the importance of viscous diffusion relative to rotation, and the thermometric properties of the fluid through its kinematic viscosity $\nu$ and thermal diffusivity $\kappa$. Since $Ra$ must be large to generate turbulence, the presence of geostrophic turbulence requires that $E$ be correspondingly smaller. This regime can be accessed by exploring the simultaneous limits $Ra \rightarrow \infty$, $E\rightarrow  0$ such that $Ro\ll1$. Unfortunately, this regime is inaccessible to both direct numerical simulations (DNS) of the Navier-Stokes equations (NSE) and laboratory investigations. To date the lowest achievable Ekman numbers are in the neighborhood of $E ={\cal{O}}(10^{-7})$, whereas an adequate exploration of geostrophic turbulence requires Ekman numbers that are much lower.
\begin{figure}
	\centering
		\includegraphics[height=0.34\textwidth]{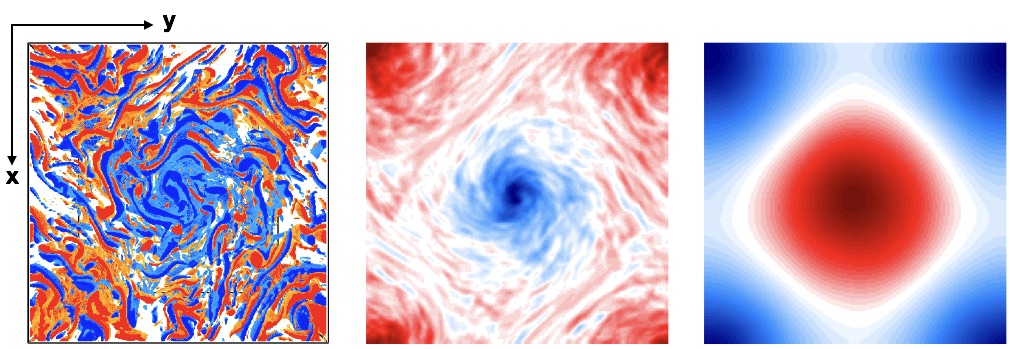}
	\caption{\small Volume render of geostrophic turbulence at $Ra E^{4/3}=90, Pr=1$. Top view of (a) total vertical vorticity $\zeta$, (b) barotropic vorticity $\left \langle {\zeta} \right \rangle$ and (c) barotropic streamfunction $\left \langle {\Psi}\right \rangle$.}
	\label{fig:Geos_vol}
	\vspace{-3ex}
\end{figure}

An alternative and fruitful approach that has recently been advanced \citep{Julien07} employs an asymptotic reformulation of the NSE for incompressible thermal convection valid in the limit $Ro\downarrow  0$ to derive a reduced system of PDEs called the nonhydrostatic quasi-geostrophic equations (NH-QGE). This reduced system filters out fast inertial waves and thin Ekman boundary layers and is therefore amenable to extensive numerical explorations.
These have been validated qualitatively by DNS studies at moderately low $Ro$ and $E$ \citep{bF14,cG14,Stellmach,plumley2016} and enabled a comprehensive mapping of the $Ra$-$Pr$ space \citep{Julien12,Rubio14}. Figure~\ref{fig:Geos_vol} illustrates volume renderings of the vorticity and streamfunction fields in the regime of geostrophic turbulence in a domain of unit aspect ratio in the horizontal. A remarkable feature of this state is the presence of a strong inverse energy cascade resulting in a box-scale condensate in the form of a vortex dipole. The feature appears to be barotropic (i.e. depth-independent) as demonstrated by, and most noticeable in, the barotropic vorticity and streamfunction fields (plots b,c). The barotropic dynamics satisfies the two-dimensional (2D) barotropic vorticity equation forced baroclinically by the underlying depth-dependent geostrophic turbulence and damped by viscosity \citep{Rubio14}. The energetics of this process can be viewed as a two-way barotropic-baroclinic interaction: the barotropic dynamics is directly forced by and extracts energy from the convective (baroclinic) dynamics. This interaction can be highly efficient in that the barotropic vortex is capable of growing to large amplitudes with little impact on the underlying geostrophic turbulence as measured by the small adjustment in convective (baroclinic) kinetic energy when the pathway to exciting the barotropic manifold is switched on at $t=0$ \citep{Rubio14}. At the same time the baroclinic fluctuations are aligned by the barotropic flow leading to a self-sustaining process.

The generation of large-scale structure in turbulent flows is primarily investigated in 2D, and focuses on the 2D Navier-Stokes equations with damping (provided by Rayleigh friction) and dissipation (provided by viscosity), and driven by externally imposed noise, usually taken to be white \citep{Smith94,bouchet09,Laurie17}. Numerical study of this system in a periodic domain with a square aspect ratio also realizes condensation into a box-scale vortex structure. When the aspect ratio becomes elongated turbulent jets oriented parallel to the short side form instead of a box-scale vortex. As shown recently \citep{Laurie17} these jets may be populated by large numbers of prominent vortices embedded in an anisotropic turbulent background state. This type of condensation process has also been examined using ideas from equilibrium statistical mechanics \citep{bouchet09,bouchet2012} which predicts a transition from a box-scale vortex to a jet state as the aspect ratio increases and the domain becomes rectangular. Both the simulations and theory find that jets are already present when the elongation is of order 10\%. However, the statistical approach describes only box-scale structures and so cannot examine the finer details of the turbulent jets it predicts. In addition, the 2D system is a driven dissipative system, and any predictions from equilibrium statistical mechanics have to be treated with caution despite the similarities between the predictions and the numerical simulations.

In the present paper we also identify a transition between a box-scale vortex dipole and jets, and also find that jets first appear when the elongation is of order 10\%. However, our system is quite different from the 2D damped noise-driven Navier-Stokes equations studied by \cite{Smith94}, \cite{bouchet09} and \cite{Laurie17} in that our equations are fully three-dimensional (3D) and the fluctuations driving the condensation process have to be determined self-consistently with the vortices or jets they produce. Thus the noise process is both anisotropic and non-white and the physics behind the condensation process necessarily differs. Our conclusion, elaborated further below, is that the condensation process is highly robust, both with respect to the physics behind the fluctuations and the substantially different nature of the governing equations themselves.

\vspace{-2ex}
\section{The Non-Hydrostatic Reduced Equations}
\label{sec:NHQGE}
A complete derivation and discussion of the NH-QGE is presented in \citet{Sprague,Julien07,Julien12}. The equations are obtained as the leading order reduction of the incompressible NSE based on a multiscale asymptotic expansion in $Ro=E^{1/3}\equiv\epsilon\ll1$ employing a small scale $L=\epsilon H\ll1$ as well as the large vertical scale $H$. For the case of stress-free upper and lower boundaries, the leading order velocity field $\boldsymbol{u}\equiv(\boldsymbol{u}_\perp,W)$ is in geostrophic balance, i.e., $ \boldsymbol{\widehat{z}} \times \boldsymbol{u}_\perp = -\nabla_\perp p$.  It follows that the horizontal velocity field is non-divergent with $\boldsymbol{u}_\perp \equiv (u,v,0)=(-\partial_{y}\Psi,\partial_{x}\Psi,0)$, where the pressure $p\equiv\Psi$ is the geostrophic streamfunction.  The reduced NH-QGE governing the motion of the fluid are
\begin{equation}\label{eq:xi}
D_t^{\bot} \zeta - \partial_Z W = \nabla^2_{\bot} \zeta \, ,
\end{equation}
\begin{equation}\label{eq:W}
D_t^{\bot} W + \partial_Z \Psi = \frac{\widetilde{Ra}}{Pr} \Theta' + \nabla^2_{\bot} W \, ,
\end{equation}
\begin{equation}\label{eq:Theta}
D_t^{\bot} \Theta' + W\partial_Z \overline{\Theta} = 
\frac{1}{Pr} \nabla^2_{\bot}  \Theta' \, ,
\end{equation}
\begin{equation}\label{eq:mTheta}
\partial_\tau \overline{\Theta}  +  \partial_Z\left ( \overline{W\Theta'}\right ) = \frac{1}{Pr} \partial_{ZZ}  \overline{\Theta} \, ,
\end{equation}
capturing, respectively, the evolution of vertical vorticity $\zeta=-\nabla^2_\perp\Psi$, vertical velocity $W$,  and temperature $\Theta = \overline{\Theta} + \epsilon  \Theta'$ at the reduced Rayleigh number $\widetilde{Ra}=Ra\, \epsilon^4$ for a given Prandtl number $Pr$. The temperature is decomposed into a mean (horizontally-averaged) component $\overline{\Theta}$ evolving on the slow timescale $\tau = \epsilon^2 t$ and a small fluctuating component $\Theta'$. Here $D_t^{\bot}\equiv\partial_t  +  \boldsymbol{u}^\prime_{\perp} \cdot \nabla_\perp$ denotes the horizontal material derivative. The system is solved with impenetrable, stress-free, fixed temperature boundary conditions:
\begin{equation}\label{eq:bdTSF}
\overline{\Theta}=1,  \ \ W=\Theta' = 0 \quad\textrm{ at} \quad  Z = 0 \, ,\qquad 
\overline{\Theta}=0,  \ \  W=\Theta' = 0 \quad\textrm{ at} \quad  Z = 1 \, .
\end{equation}

The NH-QGE are discretized in the horizontal and vertical spatial directions using a sparse Fourier-Chebyshev spectral decomposition  \citep{Watson}. 
They are then time-evolved using a third-order semi-implicit explicit Runge-Kutta scheme.

\vspace{-3ex}
\section{Results and Discussion}
\label{sec:Results}
In the following, we present results for RRBC obtained from a series of simulations of the NH-QGE system (\ref{eq:xi})-(\ref{eq:bdTSF}) performed within the geostrophic turbulence regime at $\widetilde{Ra}=90$, $Pr=1$. The horizontal aspect ratio 1:$\Gamma$ is varied from 1:1 to 1:6.  All cases exhibit similar efficiency in heat transport as measured by the Nusselt number, viz. $Nu=36.84\pm 1.97$. 

\subsection{Visualizations}
\begin{landscape}
\begin{figure} 
\begin{tabular}{lll}
\subfloat{\includegraphics[height = 0.47in]{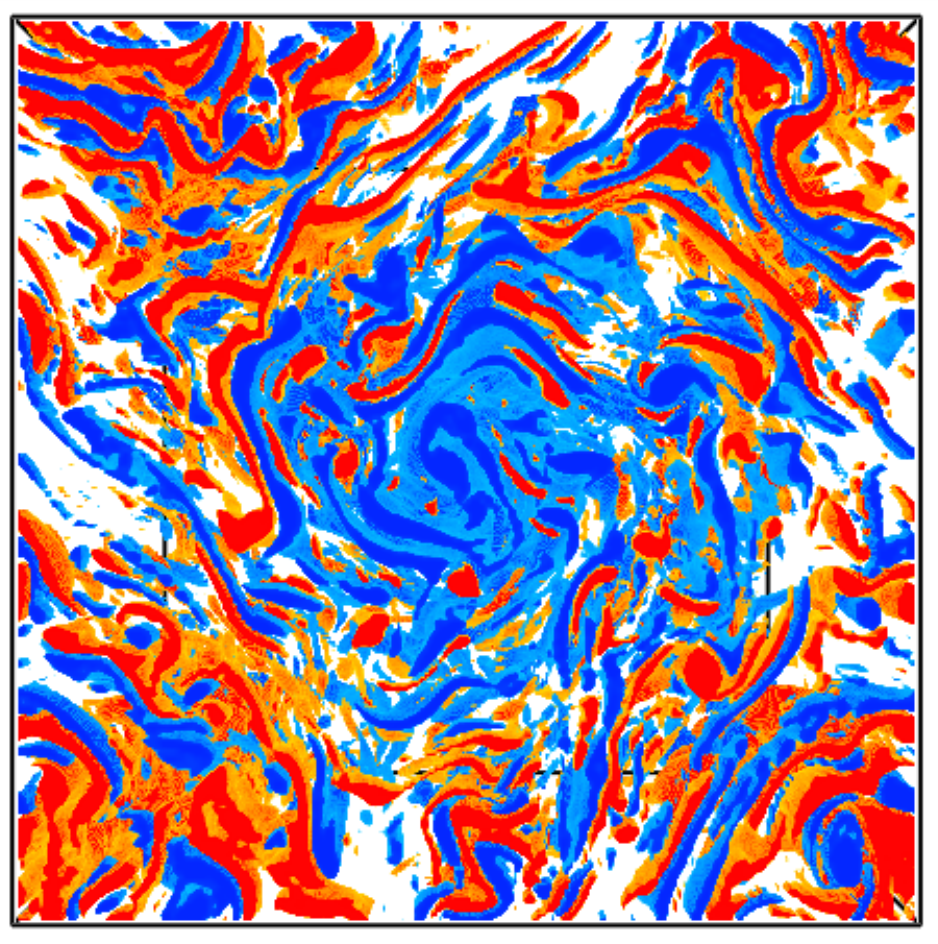}} &
\subfloat{\includegraphics[height = 0.47in]{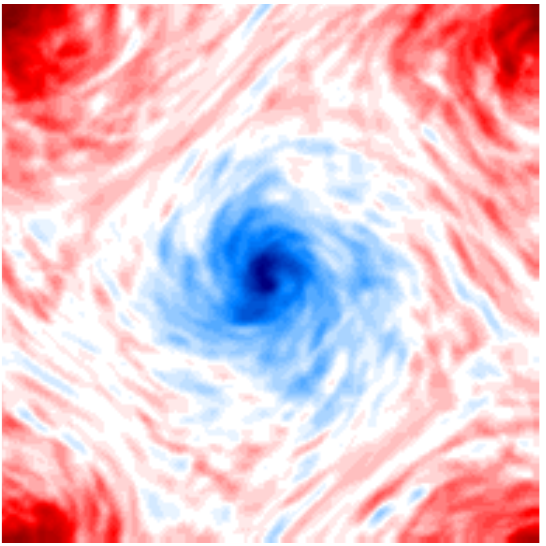}} &
\subfloat{\includegraphics[height = 0.47in]{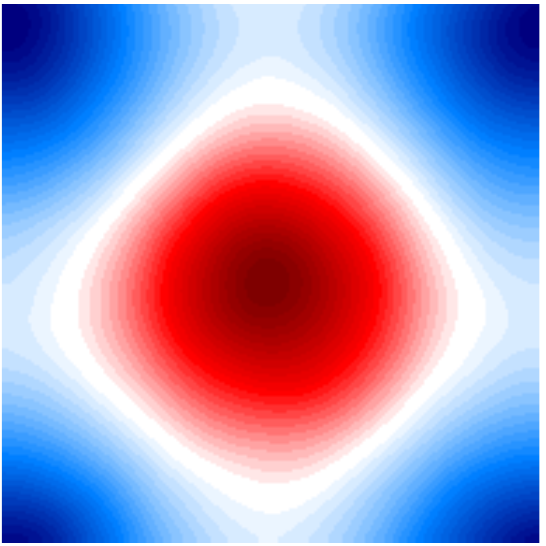}} \\
\subfloat{\includegraphics[height = 0.47in]{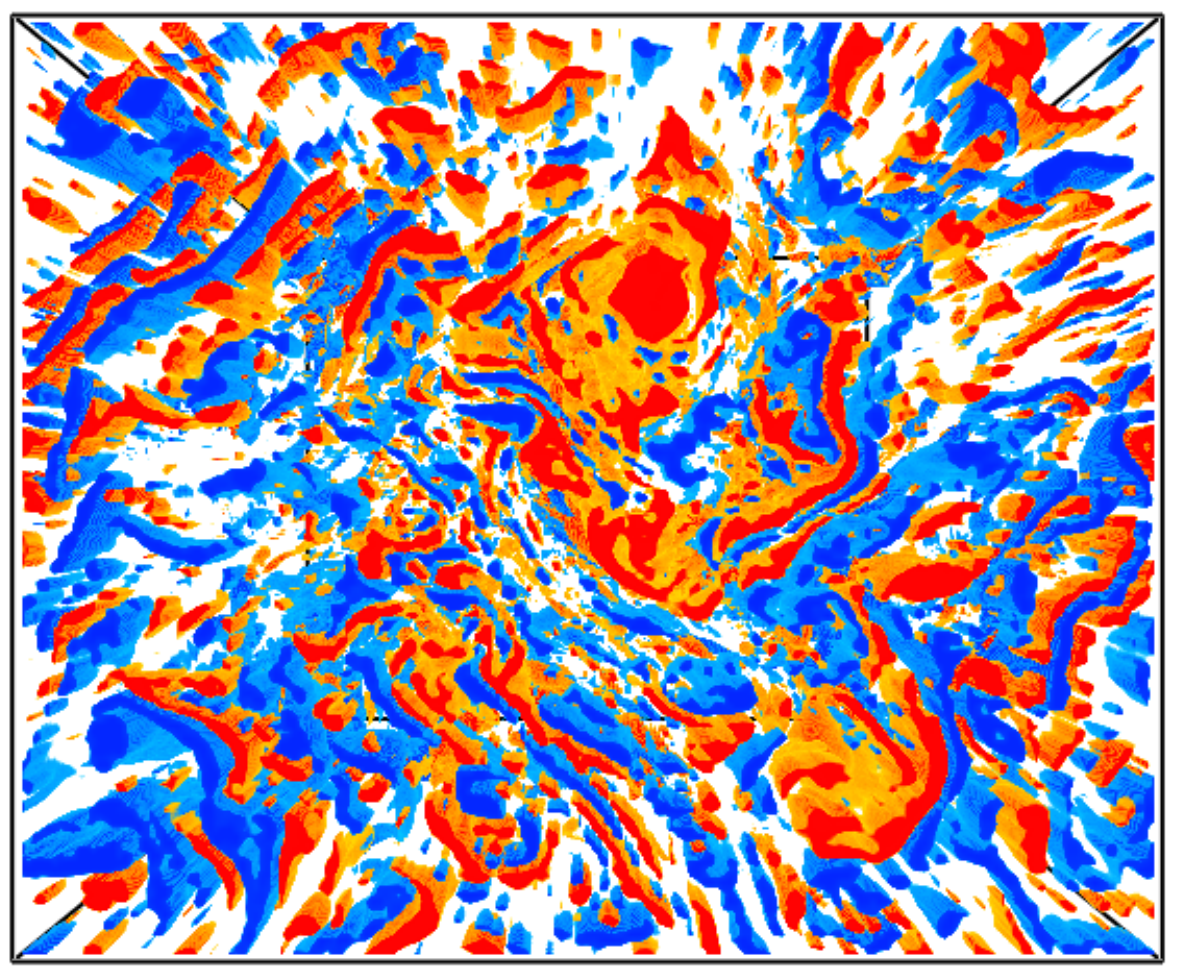}} &
\subfloat{\includegraphics[height = 0.47in]{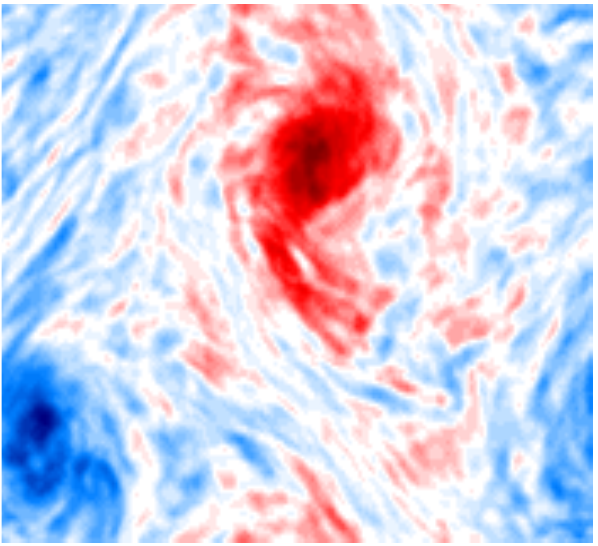}} &
\subfloat{\includegraphics[height = 0.47in]{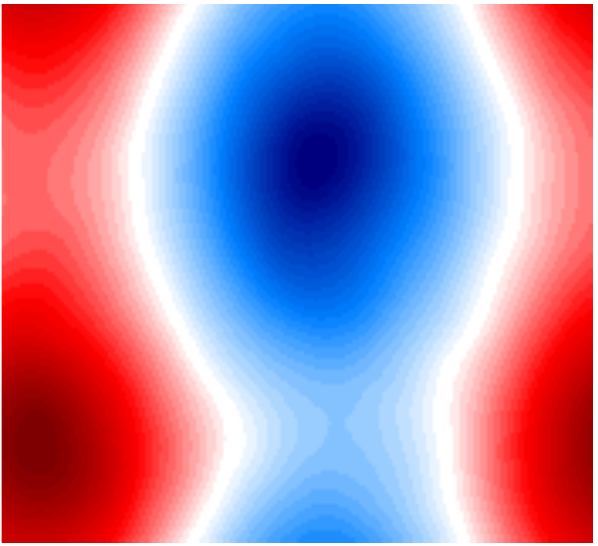}} \\
\subfloat{\includegraphics[height = 0.47in]{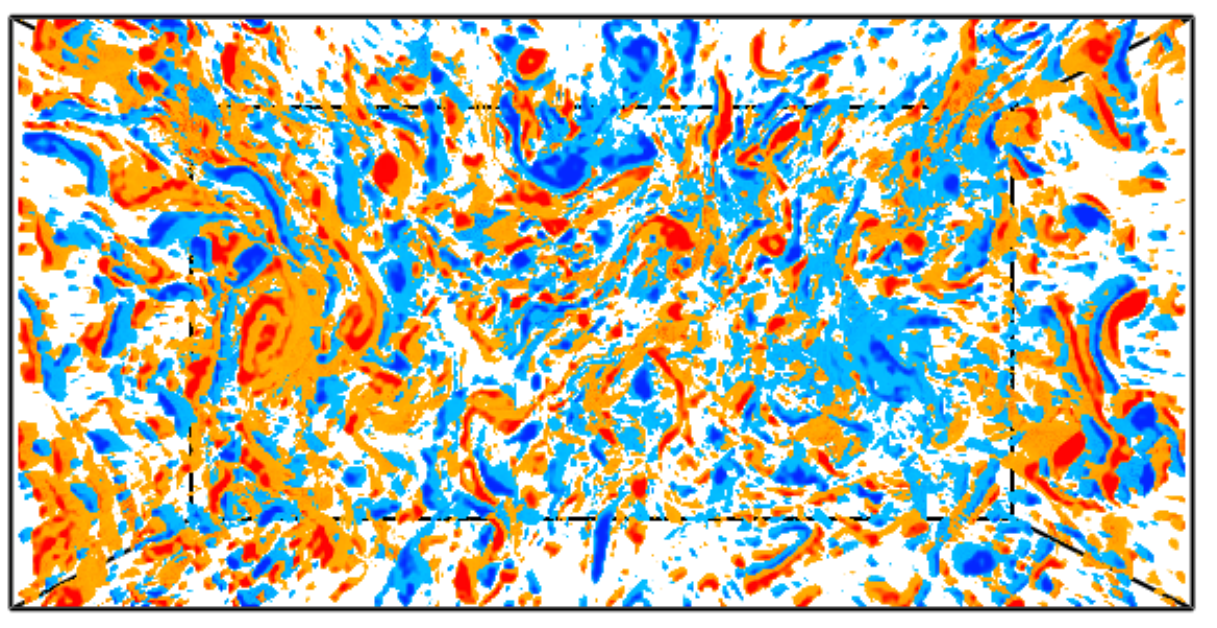}} &
\subfloat{\includegraphics[height = 0.47in]{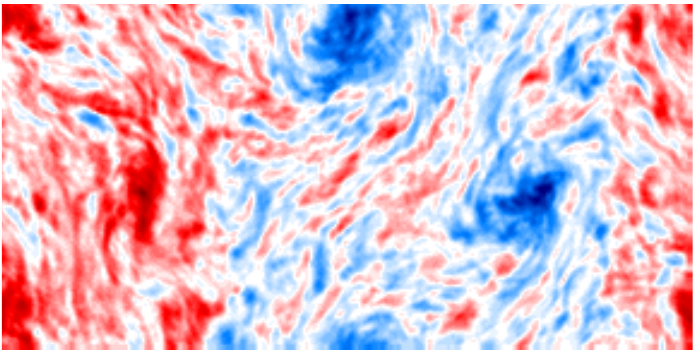}} &
\subfloat{\includegraphics[height = 0.47in]{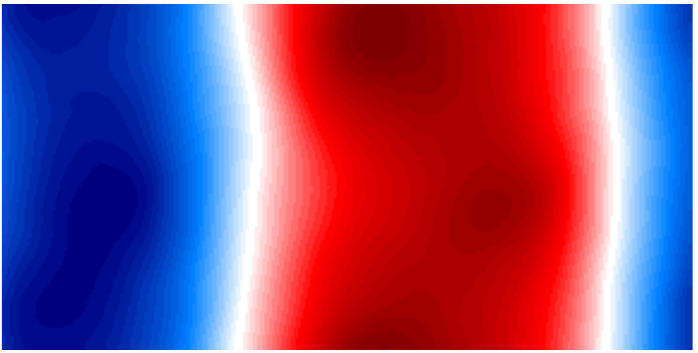}} \\
\subfloat{\includegraphics[height = 0.47in]{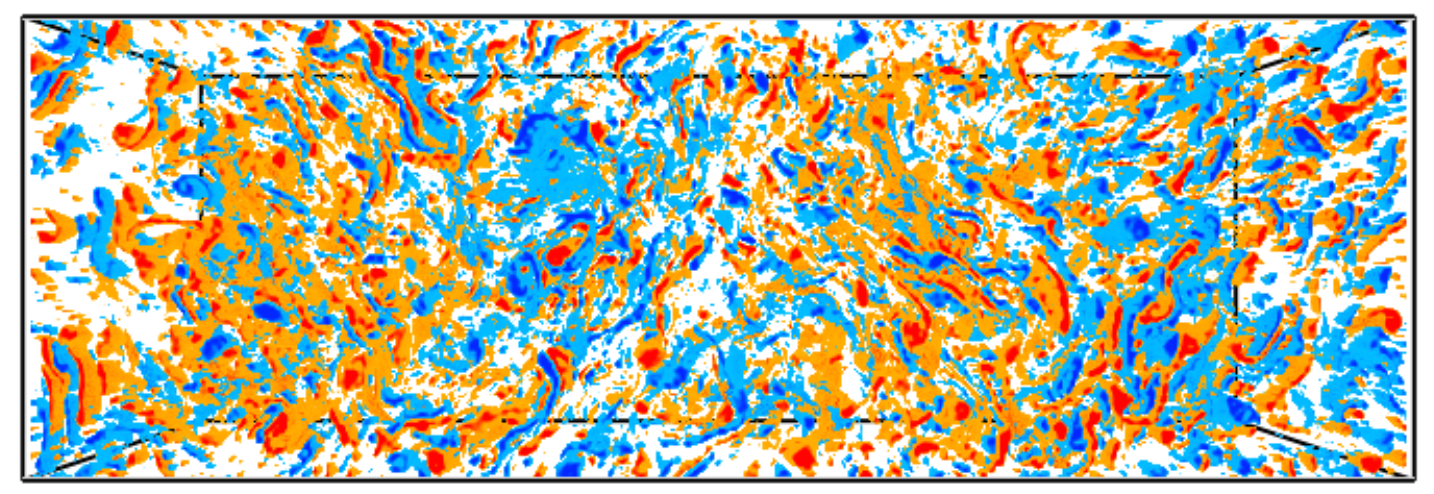}} &
\subfloat{\includegraphics[height = 0.47in]{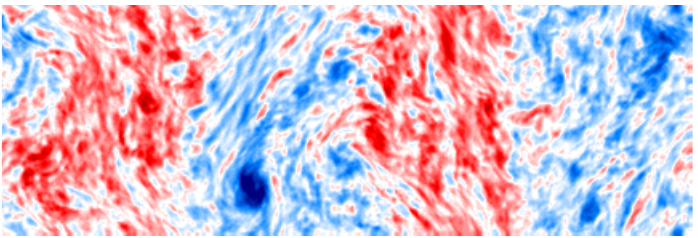}} &
\subfloat{\includegraphics[height = 0.47in]{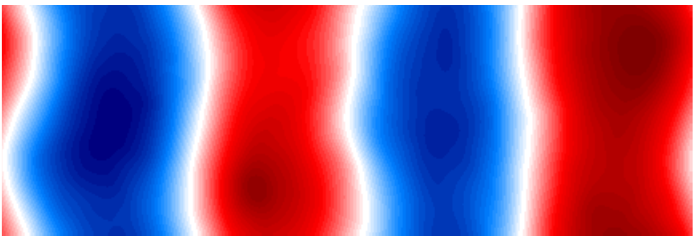}} \\
\subfloat{\includegraphics[height = 0.4896in]{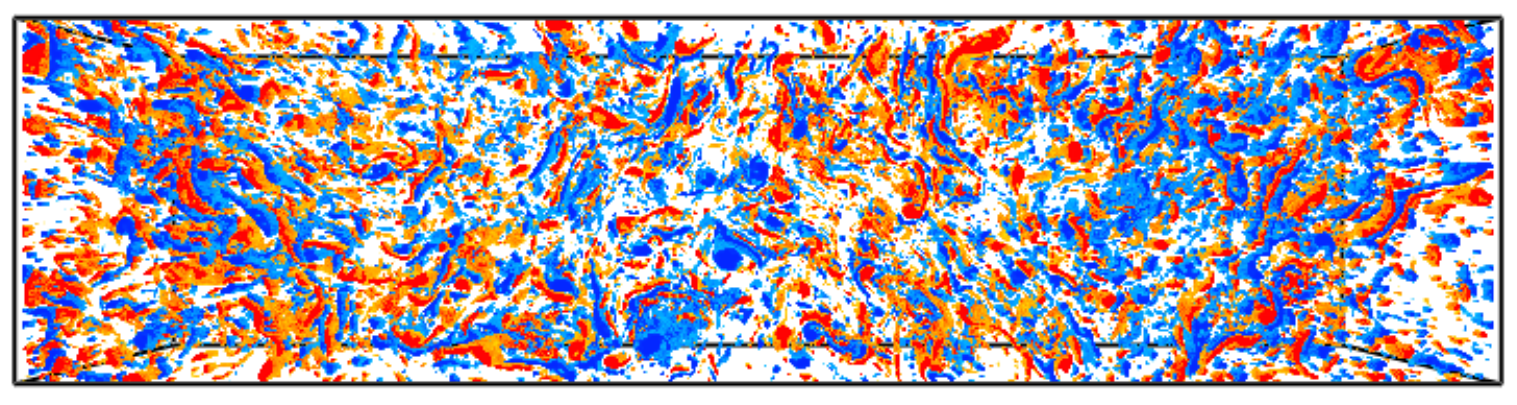}} &
\subfloat{\includegraphics[height = 0.47in]{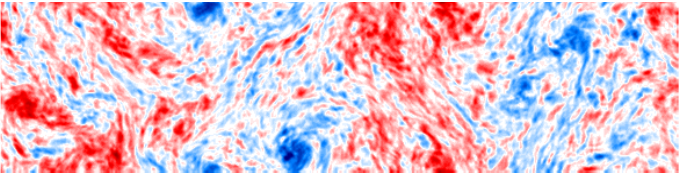}} &
\subfloat{\includegraphics[height = 0.47in]{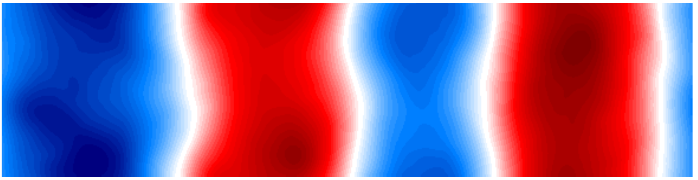}} \\
\subfloat{\includegraphics[height = 0.51in]{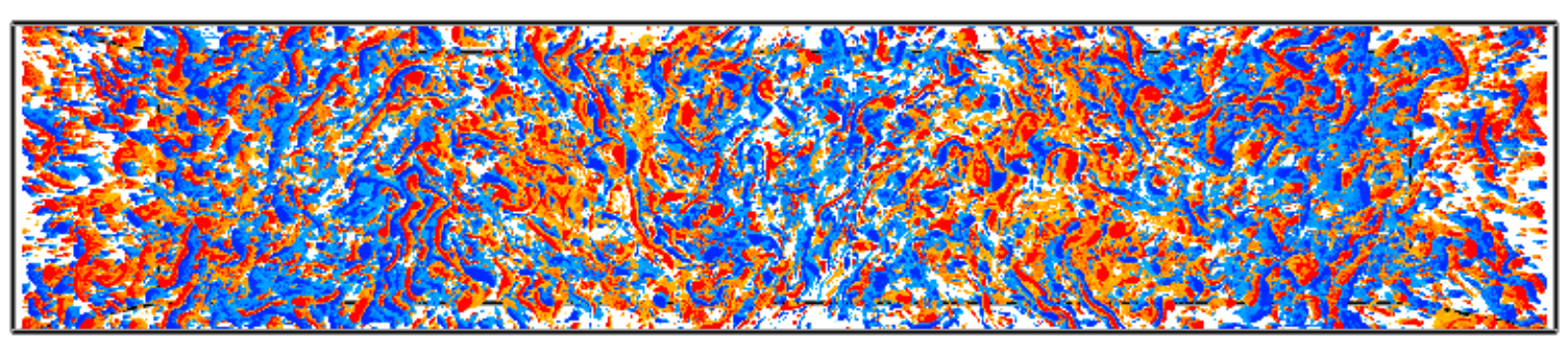}} &
\subfloat{\includegraphics[height = 0.47in]{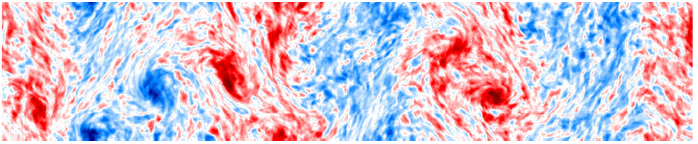}} &
\subfloat{\includegraphics[height = 0.47in]{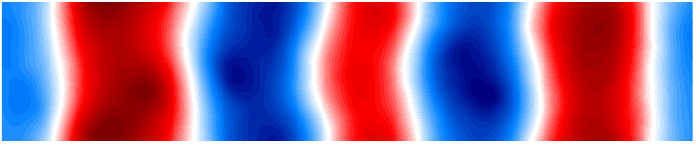}} \\ \addtocounter{subfigure}{-18}
\subfloat[$\zeta$]{\includegraphics[height = 0.4896in]{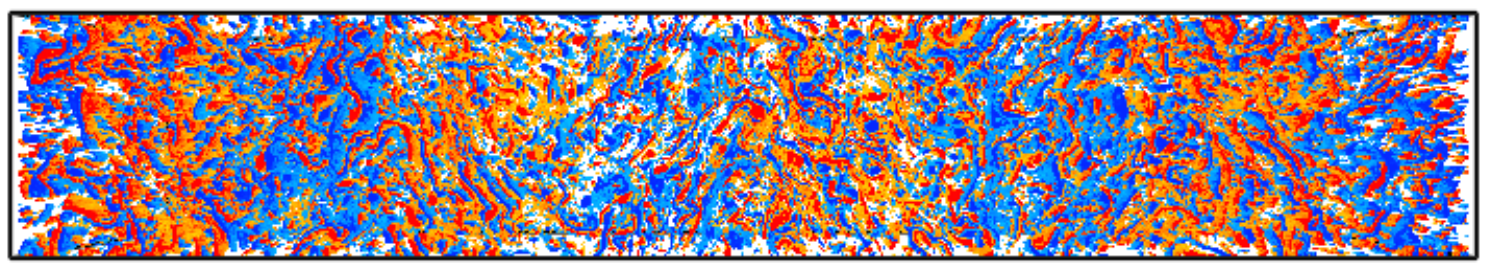}} &
\subfloat[$\left\langle\zeta\right\rangle$]{\includegraphics[height = 0.47in]{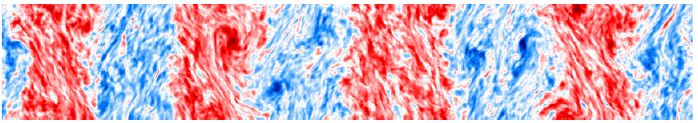}} &
\subfloat[$\left\langle\Psi\right\rangle$]{\includegraphics[height = 0.47in]{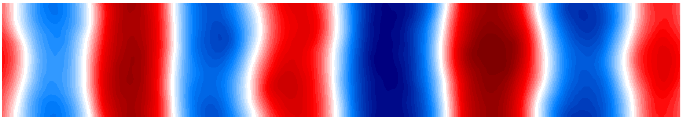}} 
\end{tabular}
\caption{{\small Top view of a suite of simulations using the NH-QGE for RRBC with impenetrable, stress-free, fixed temperature boundaries. All simulations were performed at $\widetilde{Ra}=90$, $Pr=1$. Leftmost, middle and rightmost columns display, respectively, the total vorticity $\zeta$, barotropic vorticity $\left\langle{\zeta}\right\rangle$, and barotropic streamfunction $\left\langle{\Psi}\right\rangle$. In units of a fixed $x$-horizontal box scale $L_x=10 L_c$ (where $L_c=4.82 E^{1/3} H$ is the critical wavelength for convection) the domain aspect ratio $(1:\Gamma)L_x$ increases from 1:1 (upper row), 1:1.1 (second row), and incrementally from 1:2 to 1:6 (remaining rows). Note that $x$ is in the vertical direction with $y$ plotted horizontally. Evident is the transition to banded flows of alternating cyclonicity when $\Gamma=2$ and a doubling, tripling and quadrupling of the band structure at $\Gamma=3$, $5$ and $6$, respectively. Second row ($\Gamma=1.1$) shows that the transition to banded flow occurs upon even small departure from isotropy. Resolution in the $x,y,Z$ directions is $144 \times 144 \Gamma\times 192$. }}
\label{fig:allapsectratios}
\end{figure}
\end{landscape}

Depicted in Fig.~\ref{fig:allapsectratios} are the top views of volume renderings of the total vertical vorticity $\zeta$ (left column), barotropic vorticity  $\left\langle{\zeta}\right\rangle$ (middle column), and the barotropic streamfunction $\left\langle{\Psi}\right\rangle$ (right column). Here $\langle\cdots\rangle$ indicates average in the vertical. The barotropic vortex dipole present at aspect ratio 1:1 (top row) is replaced by a state of approximately parallel (i.e., banded) flow consisting of an alternating sequence of cyclonic and anti-cyclonic vortical bands. This transition first occurs at approximately aspect ratio 1:1.1 (second row) and is most prominently revealed in the rendering of the barotropic streamfunction; $\left\langle{\zeta}\right\rangle$ exhibits greater spatial complexity due to higher spatial derivatives. We see that strong vortical eddies persist within a band of given cyclonicity and that small-scale geostrophic turbulence is globally advected and organized by the banded large-scale flow. {The latter resembles the small-scale filamentary structures in Fig~\ref{fig:Geos_vol}a.} From a more global perspective, the aspect ratio $\Gamma$ provides a selection mechanism for the number of alternating bands or jets. It can be seen that an increase from $\Gamma=2$ to $\Gamma=3$ results in a doubling of the number alternating bands from one to two. The latter persists for $\Gamma=4$ before losing stability to a state of three alternating bands at aspect ratio $\Gamma=5$ and then four at $\Gamma=6$.

\begin{figure} 
\centering
\begin{tabular}{rl}
\vspace{-1ex}
\subfloat[$\langle u \rangle \  (1:2)$]{\includegraphics[height = 0.6in]{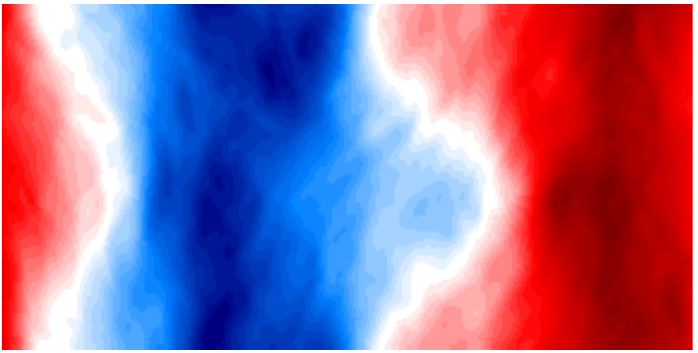}} & \hspace{1.5em}
\subfloat[$\langle v\rangle \  (1:2)$]{\includegraphics[height = 0.6in]{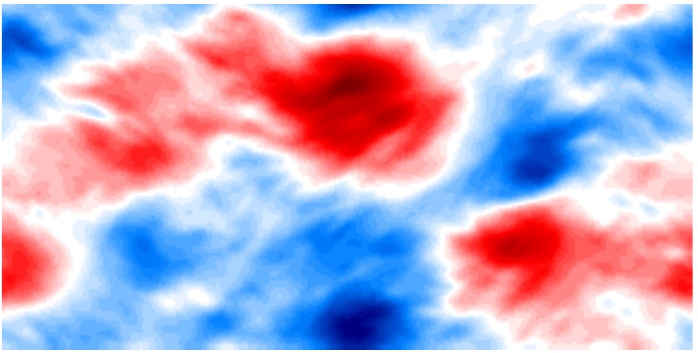}} \\
\vspace{-1ex}
\subfloat[$\overline{\langle u \rangle}^x (y)\  (1:2)$]{\includegraphics[height = 0.82in]{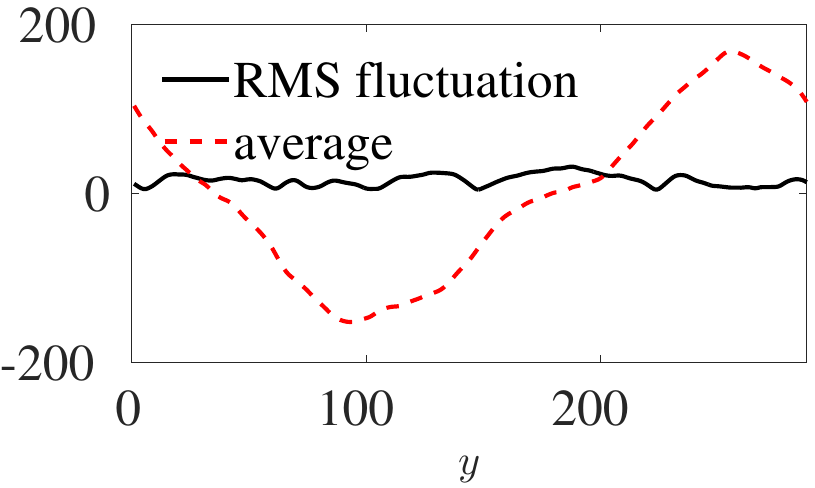}} &\hspace{.1em}
\subfloat[$\overline{\langle v\rangle }^y (x)\  (1:2)$]{\includegraphics[height = 0.81in]{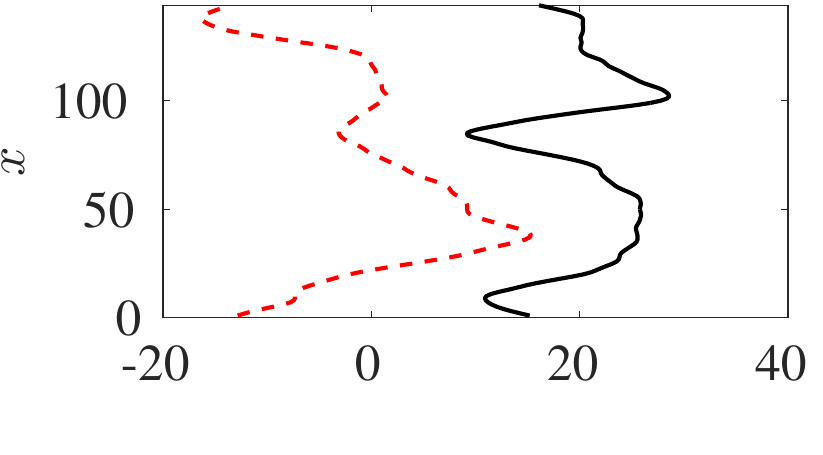}}
\end{tabular}
\\
\begin{tabular}{rl}
\vspace{-1ex}
\subfloat[$\langle u \rangle \  (1:5)$]{\includegraphics[height = 0.5in]{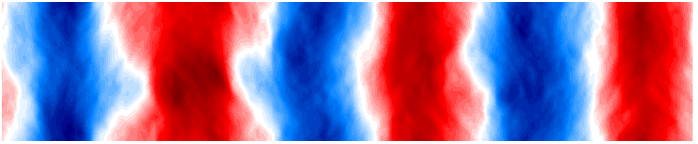}} & \hspace{1.5em}
\subfloat[$\langle v \rangle \  (1:5)$]{\includegraphics[height = 0.5in]{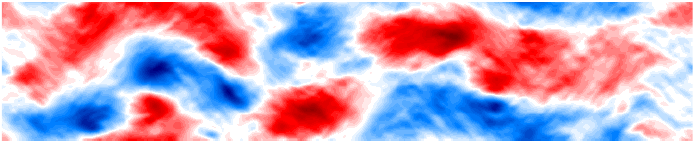}} \\
\vspace{-1ex}
\subfloat[$\overline{\langle u \rangle}^x (y)\  (1:5)$]{\includegraphics[height = .8in]{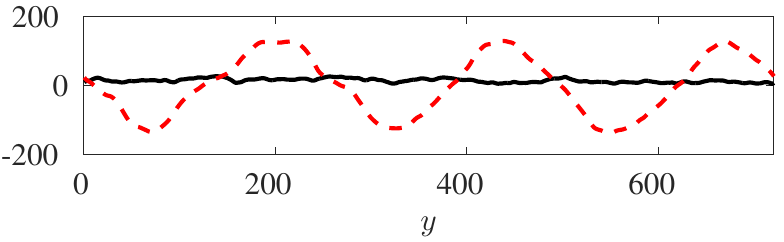}} & 
\subfloat[$\overline{\langle v\rangle }^y (x)\  (1:5)$]{\includegraphics[height = 0.78in]{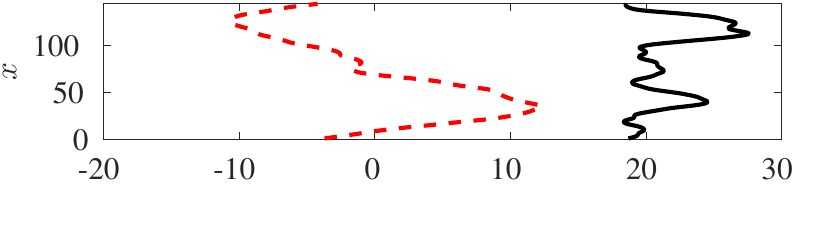}}
\end{tabular}
\\
\begin{tabular}{rl}
\vspace{-1ex}
\subfloat[$\overline{\langle u \rangle }^x vs\ t$]{\includegraphics[height = 0.7in]{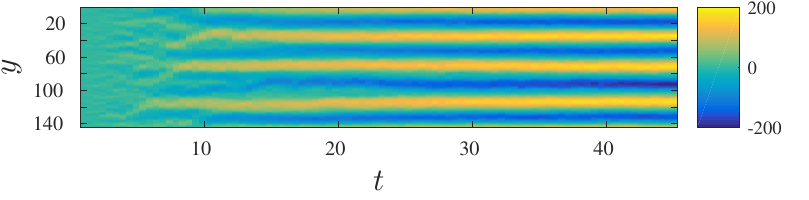}} &
\subfloat[$\overline{\langle v\rangle }^y vs\ t$]{\includegraphics[height = 0.7in]{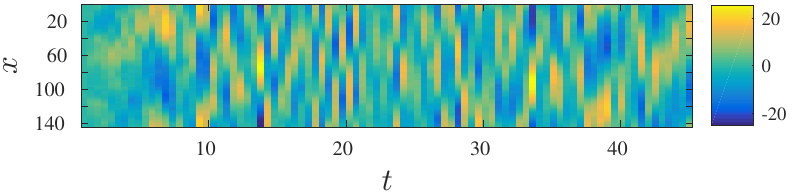}} 
\end{tabular}
\vspace{-2ex}
\caption{{\small Parallel and tranverse barotropic velocity fields $\langle u \rangle $ (left column) and $\langle v\rangle $ (right column).  Aspect ratios 1:2 and 1:5 are depicted in the top two and middle two rows, respectively.  
Here $\overline{ f }^d $ indicates averaging of $f$ in the $d$ direction. The average and RMS fluctuation of $\langle u \rangle$ in the parallel direction, i.e., averaging in $x$ direction, reveal a Kolmogorov-like velocity profile and of $\langle v \rangle$ in the transverse direction reveal the presence of a meandering jet {(red dashed lines). The amplitude of the RMS fluctuations about these profiles is shown in solid black lines.}  The bottom row shows Hovm\"oller diagrams of (i) $\overline{\langle u \rangle}^x$  and (j) $\overline{\langle v \rangle}^y$ for aspect ratio 1:5 at time intervals $\Delta t=0.5$.  }}
\label{fig:lateraljet}
\end{figure}

Inspection of the horizontal velocity fields offers another viewpoint for interpreting the large-scale barotropic structure (Fig.~\ref{fig:lateraljet}). In the following, we refer to the {velocity components $u=-\partial_y\Psi$ and}
 {$v=\partial_x\Psi$ as parallel and transverse, respectively. The parallel velocity field (left column) clearly shows a state resembling Kolmogorov flow, i.e., alternating bands of unidirectional but oppositely directed turbulent flow (plots a,e). Averaging along the $x$-direction reveals a jet structure that is sawtooth in shape in the $y$ or transverse direction (dashed lines, plots c,g), with small RMS fluctuations (solid lines, plots c,g). This unidirectional jet structure is found at all sufficiently non-square aspect ratios
 (Fig.~\ref{fig:allapsectratios}).  Observations of $v$ (right column) indicate coexistence with a
weak meandering transverse jet (plots b,f). Averaging in the $y$-direction (dashed lines, plots d,h) shows that the transverse mean flow is substantially weaker than the parallel flow (by a factor of approximately ten). RMS fluctuations about this profile are large 
}

\noindent
 {indicating that the transverse mean flow is highly oscillatory (solid lines, plots d,h). The lowest row of Fig.~\ref{fig:lateraljet} illustrates the persistence of the parallel and transverse flows with time through Hovm\"oller diagrams. Here we see that after an initial transient stable parallel jets are formed while the antisymmetric transverse flow exhibits random switching.}

\begin{figure}
	\centering
	      \subfloat[]{\includegraphics[width=0.38\textwidth]{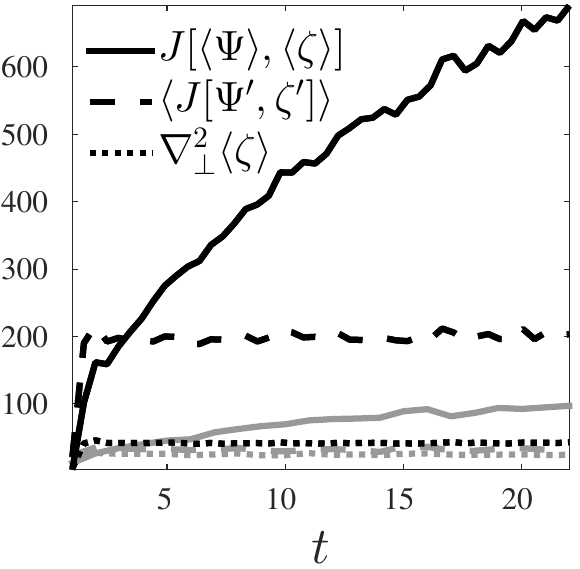}} \hspace{3em}
		  		  \subfloat[]{\includegraphics[width=0.4\textwidth]{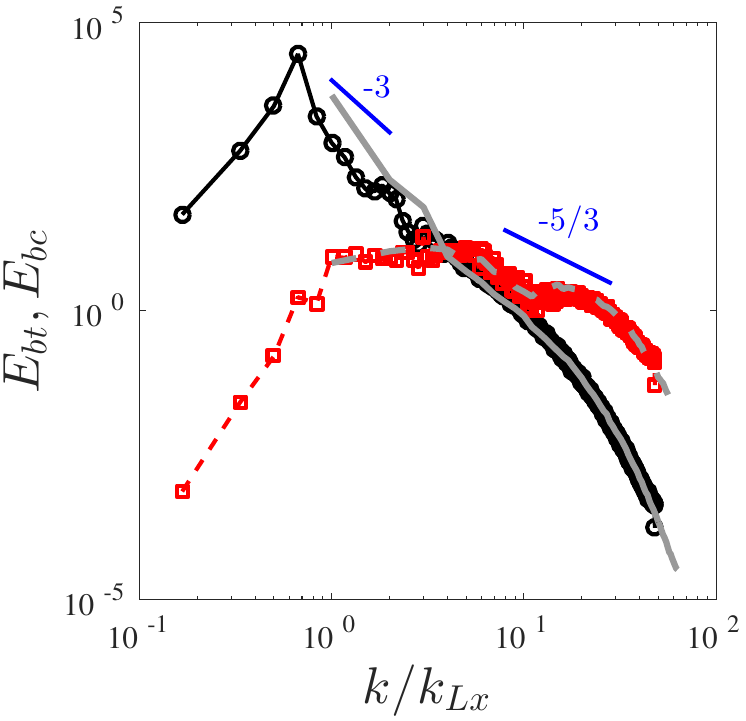}}
	          \caption{\small{(a) Term by term decomposition of the barotropic vorticity equation (\ref{eq:bxi}) showing unbounded growth in the barotropic mode for  $\Gamma=6$ (black) and $\Gamma=1$ (gray). (b) One dimensional barotropic (solid) and baroclinic (dashed) kinetic energy spectra as a function of  $k/k_{Lx}$ for $\Gamma=6$ with the corresponding results for $\Gamma=1$ in gray. $k_{Lx}=2\pi/L_x$ denotes the box wavenumber in the $x$ direction. Peak energy scales $k/k_{Lx}=4/6$ for $\Gamma=6$ and  $k/k_{Lx}=1$ for $\Gamma=1$  characterize the  jet scale.  }}
	\label{fig:energy}
	\vspace{-1ex}
\end{figure}

\subsection{Energetics}

In the absence of dissipation, and like the incompressible NSE, all QG systems conserve the volume-averaged energy ${\cal{E}}$ and the pointwise potential vorticity ${\cal{PV}}$. For the NH-QGE (\ref{eq:xi})-(\ref{eq:mTheta}) these are given by
\begin{equation}
{\cal{E}} =  \left \langle \left ( \frac{1}{2} \overline{\vert \nabla_\perp\Psi \vert^2   + W^2} \right ) + \frac{\widetilde{Ra}}{Pr}\left ( \overline{ \Theta^{\prime 2} }
+\epsilon^{-2} \left (  \overline{\Theta} - z \right )^2  \right ) \right \rangle,
\end{equation}
\begin{equation}
\label{eq:PV}
{\cal{PV}} = \zeta - J\left [W, \frac{\Theta'}{\partial_{Z} \overline{\Theta}} \right ]  +\partial_{Z}\left ( \frac{\Theta'}{\partial_{Z} \overline{\Theta}}\right )\,,
\end{equation}
where   
$J[f,g]=\partial_x f \partial_y g - \partial_y f \partial_x g$. As a consequence of strong vertical motions, with $\vert W\vert\sim\vert\boldsymbol{u}_\perp\vert$, ${\cal{E}}$ and ${\cal{PV}}$ are not solely functionals of the geostrophic streamfunction $\Psi$: the second and third terms in equation (\ref{eq:PV}) represent ageostrophic baroclinic contributions to ${\cal{PV}}$. Sole functional dependence on $\Psi$ may be recovered from the NH-QGE system only in the limit of strong stratification where $\partial_{Z}\overline{\Theta}\rightarrow\epsilon^{-1}$, $\partial_{Z}\rightarrow\epsilon^{-1}$ and $W\rightarrow\epsilon$.  Here, hydrostatic balance in equation (\ref{eq:W}) implies $\Theta'\rightarrow (Pr/\widetilde{Ra}) \partial_Z \Psi$ and classical H-QGE is recovered. In this case, the volume-averaged potential enstrophy $\left \langle \overline{{\cal{PV}}^2}\right \rangle$ becomes the second conserved quantity required to guarantee a dual cascade. 

As already noted, however, the dynamics within the barotropic subspace provides an alternative pathway for an inverse cascade. Depth-averaging the vertical vorticity equation (\ref{eq:xi}) gives the barotropic vorticity equation (BVE)
\begin{equation}\label{eq:bxi}
\partial_t \langle \zeta \rangle+   J[ \langle \Psi \rangle, \langle \zeta \rangle] = -  \langle J[  \Psi^\prime ,  \zeta^\prime ]  \rangle
 + \nabla^2_\perp \langle \zeta \rangle \, ,
\end{equation}
indicating that the material growth of $\langle \zeta \rangle$ depends on the net balance between the two terms on the right-hand side, i.e., between the baroclinic forcing and barotropic viscous dissipation.  In the absence of forcing and damping, the conserved quantities are the volume-averaged barotropic energy and enstrophy
\begin{equation}
{\cal{E}}_{bt} = \frac{1}{2}  \left\langle \overline{\vert {\nabla_\perp}\left \langle \Psi  \right \rangle \vert^2} \right\rangle,\qquad
{\cal{Z}}_{bt} = \overline{ \left  \langle \zeta  \right \rangle^2}=
\left \langle\overline{\left (  \nabla_\perp^2  \langle \Psi   \rangle\right )^2} \right\rangle .
\end{equation}
Both are sole functionals of $\langle \Psi \rangle$ suggesting a dual cascade. For all aspect ratios we find that the evolution of the volume-averaged kinetic energy is similar to the isotropic case. Figure~\ref{fig:energy}a illustrates the decomposition of the right side of the BVE (\ref{eq:bxi}) as a function of time for $\Gamma=1$ and $\Gamma=6$. The figure clearly demonstrates that in both cases convective forcing $\langle J[ \Psi', \zeta' ] \rangle$ and damping $\nabla^2_\perp \langle \zeta\rangle$ are saturated but unbalanced, resulting in unbounded growth of the large-scale barotropic mode. However, despite this similarity, notable distinctions exist in the energetics. Figure~\ref{fig:energy}b shows the corresponding kinetic energy spectra as a function of the renormalized horizontal wavenumber $k/k_{Lx}$ where $k=\sqrt{k_x^2+k_y^2}$ and $k_{Lx}=2\pi/L_x$  
associated with the box dimension $L_x$. Here,
\begin{equation}
E(k) =\int_0^{2\pi} E(\mathbf{k}) k d\phi_k
\end{equation}
where in polar representation $k=\vert \mathbf{k} \vert$ and $\tan \phi_k = k_y/k_x$. The spectra have also been decomposed into barotropic (bt) and baroclinic (bc) components. For both $\Gamma=1$ and $\Gamma=6$ (and indeed all intermediate cases), the barotropic signature (solid curves) exhibits a steep power law with $E_{bt}(k)\sim k^{-3}$ while the baroclinic signature (dashed curves) gains dominance at higher wavenumber and exhibits a shallower instantaneous power $E_{bc}(k)\sim k^{-5/3}$ \citep{Rubio14}. 
The total kinetic energy spectrum $E_{bt}+E_{bc}$ exhibits a steep to shallow transition in the power law exponents, a result reminiscent of the Nastrom-Gage spectrum observed in atmospheric and oceanic measurements \citep{nastrom}. In the present RRBC case, however, the $k^{-3^+}$  barotropic spectrum is a consequence of the large-scale condensate \citep{Smith99}. For $\Gamma=1$ (grey curve), we observe that the most energetic barotropic scale is $k/k_{Lx}=1$  indicating that the large-scale condensate (the vortex dipole) has reached the box scale, the largest scale possible. For anisotropic aspect ratios the most energetic scale is that associated with the unidirectional jet occurring at a scale intermediate to the box dimensions, i.e. $k_{Ly} < k< k_{Lx}$. For $\Gamma=6$, where four jets are observed this occurs at $k/k_{Lx}=4/6$ (see solid line, Fig.~\ref{fig:energy}b) with a steep decline in power from its peak to the largest box scale $2\pi/k_{Ly}$. These larger scales are associated with the weaker meandering transverse jet (see Fig.~\ref{fig:lateraljet}b,f). For $2\le\Gamma\le6$ we observe maximal power in the barotropic energy spectra at $1/2 \le k/k_{Lx}\le 4/6$, or equivalently, $3/2 \le L/{L_x} \le 2$.
 
\begin{figure}
	\centering
	        \subfloat[$1:1$]{\includegraphics[width=0.32\textwidth]{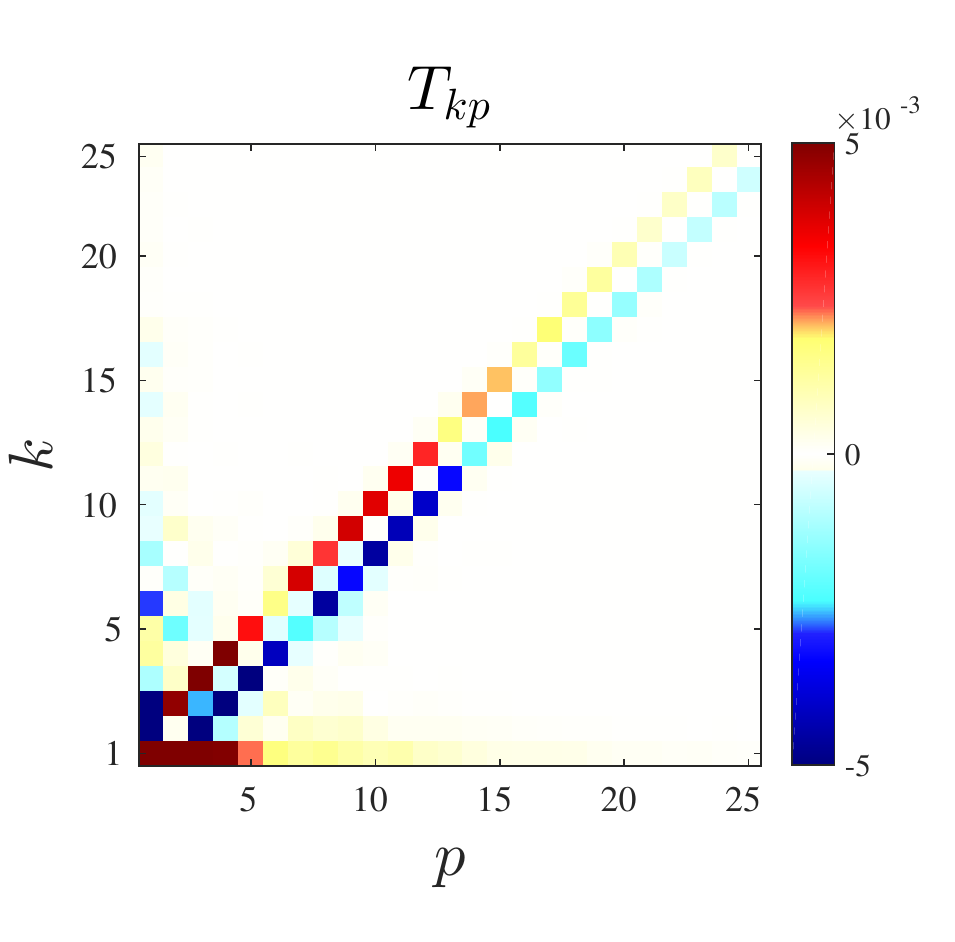}}
		  \subfloat[$1:3 $]{\includegraphics[width=0.32\textwidth]{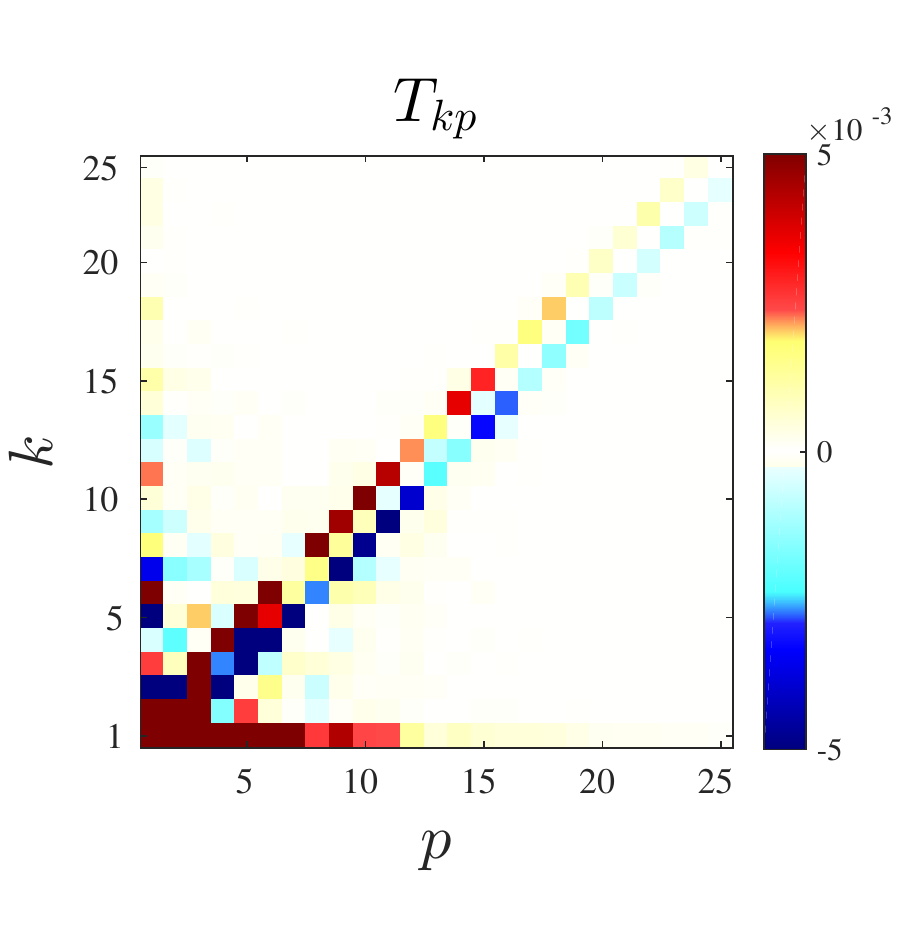}}
		  \subfloat[$1:6$]{\includegraphics[width=0.32\textwidth]{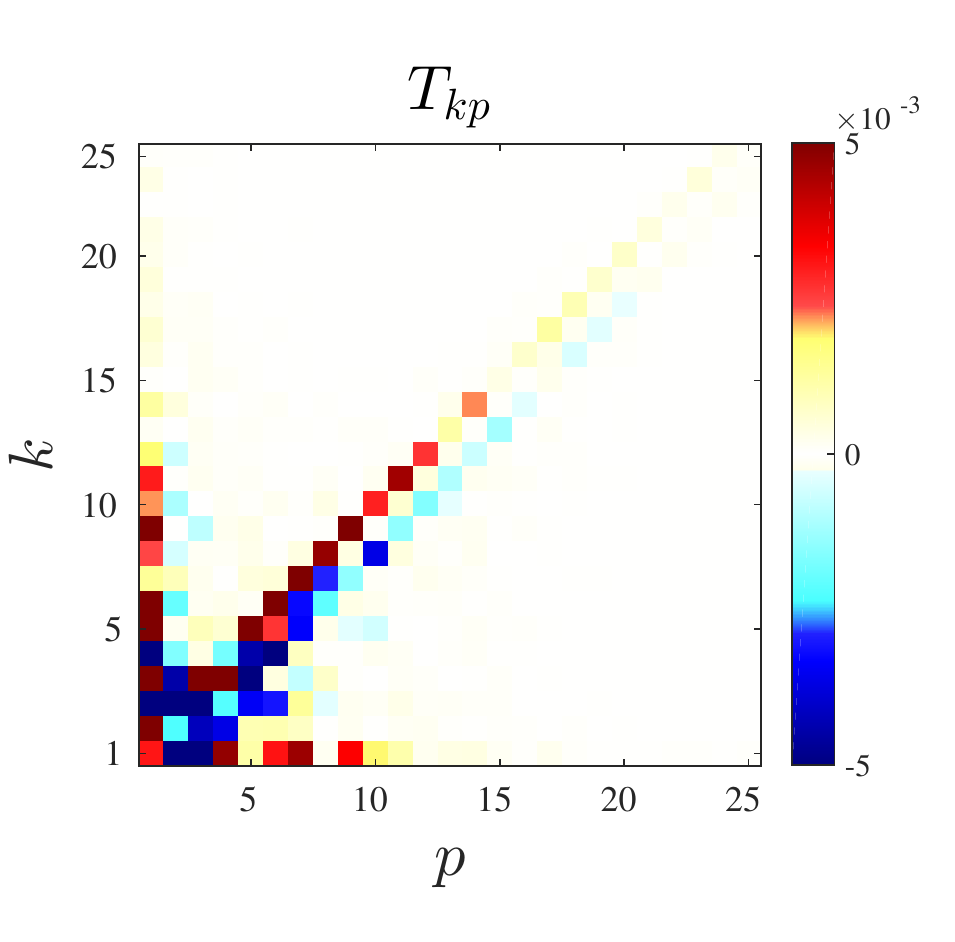}} 
	        \caption{\small{Spectral transfer maps of barotropic self-interaction for  aspect ratio (a) 1:1 (b) 1:3 and (c) 1:6, showing how energy is transferred from wavenumbers $p$ to wavenumbers $k$. The results in (b,c) have been coarse-grained to manage the non-integer wavenumbers arising from the anisotropy of the domain.   }}
	\label{fig:energymap}
	\vspace{-1ex}
\end{figure}

The flow of energy associated with the generation of large-scale dynamics in the barotropic subspace can be determined from the nonlinear advection term in (\ref{eq:bxi}). Accordingly, we detail how power is transferred to horizontal wavenumber $\mathbf{k}$ through triadic interactions involving wavenumbers $\mathbf{p}$ and $\mathbf{q}$ such that $\mathbf{p}+ \mathbf{q}+\mathbf{k}=\mathbf{0}$. We define the transfer functions
\begin{eqnarray}
T_{\mathbf{kpq}}&=&b_{\mathbf{pq}} {\rm Re} [\langle \hat \Psi_\mathbf{k} \rangle\langle \hat \Psi_\mathbf{p} \rangle\langle \hat \Psi_\mathbf{q} \rangle],\\
b_{\mathbf{pq}}&=&b_{\mathbf{qp}}=\frac{1}{2} \left ( p^2 - q^2\right ) \ \left (  p_x q_y - p_y q_x\right ),
\end{eqnarray}
where ${\rm Re}$ denotes the real part and $\delta_{\mathbf{p}+\mathbf{p}+\mathbf{p},\mathbf{0}}$ is the Kronecker delta function. Anisotropy is handled by replicating a $1:\Gamma$ barotropic field $\Gamma$ times to form a periodic square of size $(\Gamma L_x)^2$. Owing to this periodic extension, the power map of a 2D Fourier transform of the barotropic field contains the non-integer wavenumber array $(k_x,k_y)=(2\pi/ L_x ) (i/\Gamma,j/\Gamma)$ with $i,j=0,1,\dots,N_x\Gamma$. The array is sparse owing to zero row entries corresponding to wavenumbers $k_x$ with non-integer values $i/\Gamma$ that do not fulfill the periodicity of the $(0,L_x)$ domain. To avoid the impact of these zero entries when averaging over annular or spherical shells in wavenumber space we invoke coarse-graining by averaging over blocks of $\Gamma^2$ entries: the block associated with integer wavelengths $(2\pi/ L_x )(p,q)$ is indexed by
\begin{eqnarray}
 i_x = p-\frac{i-1}{\Gamma},\  i=1,\Gamma;\qquad 
 i_y = q-\frac{j-1}{\Gamma},\  j=1,\Gamma.
 \label{eqn:wave}
\end{eqnarray}
The result of coarse-graining is a power map array that contains integer wavenumbers $(k_x,k_y)=(2\pi/ L_x ) (p,q)$ with $p,q=0,1,\dots,N_x$. The coarse-grained transfer map 
\begin{eqnarray}
T_{kp} = \int k d\phi_k \int p d\phi_p \sum_{\mathbf{q}} T_{\mathbf{kpq}}
\end{eqnarray}
details the transfer of energy from wavenumber $p$ to $k$. Figure~\ref{fig:energymap} illustrates three cases: $\Gamma=1$ and the coarse-grained transfer maps for $\Gamma=3$ and $6$. The power signature in the super- and sub-off-diagonal lines in all barotropic self-interaction $T_{kp}$ maps indicates the existence of a forward or direct cascade, i.e., direct transferred of spectral power from low to high $k$ at constant wavenumber $p$. The nonlocal inverse cascade occurs for $p\gg k$ and corresponds to the direct transfer of power from the high $p$ wavenumbers to $k \approx 1$. Likewise, when $p\ll k$, energy is extracted from high $k$ wavenumbers and transferred to $p \approx 1$. Recall that for $\Gamma > 1$ the wavenumber $(p,k)=(1,1)$ associated with the large-scale structure is associated with subharmonic wavenumbers identified in (\ref{eqn:wave}) with $p,q=1$.

{ 
 \begin{figure}
 \centering
\begin{tabular}{ll}
\subfloat{\includegraphics[width=0.4\textwidth]{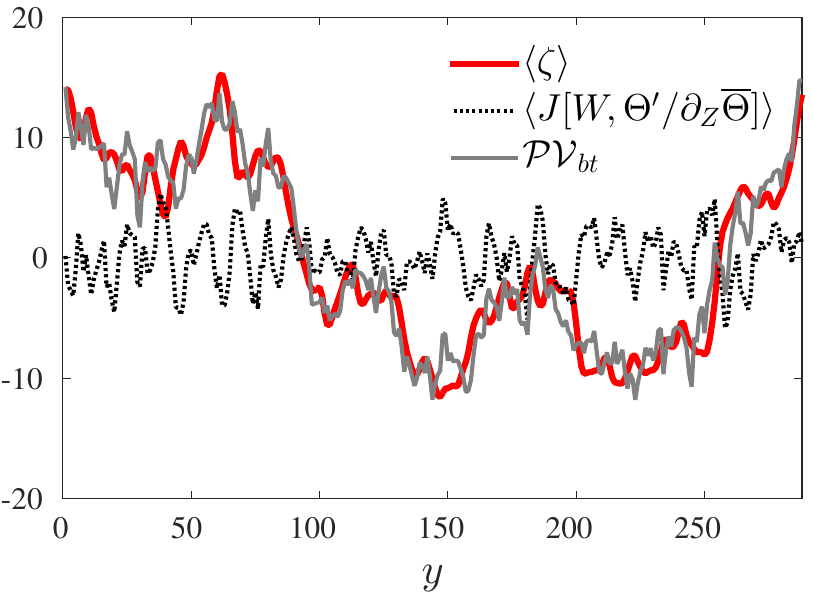}} & \hspace{1em}
\subfloat{\includegraphics[width=0.4\textwidth]{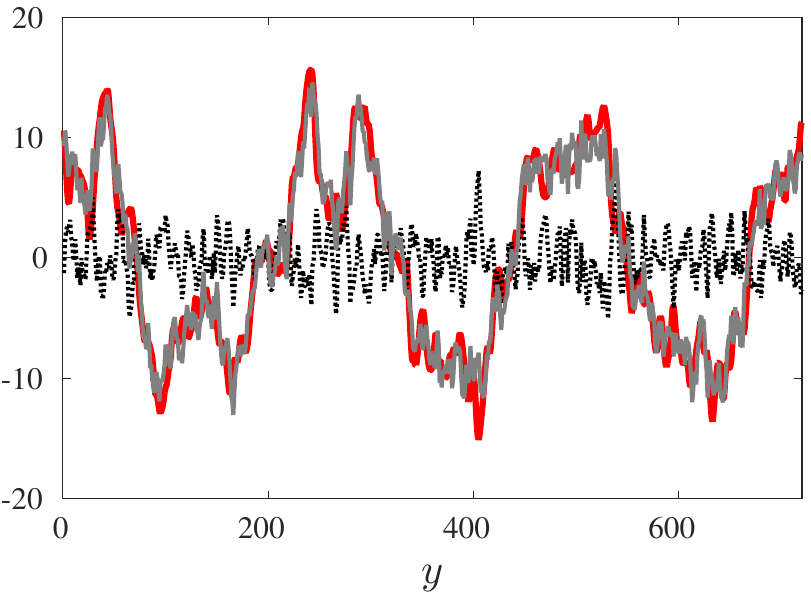}} 
\end{tabular}
	\caption{\small Barotropic potential vorticity ${\cal PV}_{bt}$ and its decomposition for aspect ratios 1:2 (left) and 1:5 (right). The $\langle \zeta \rangle$ term dominates, suggesting the linear relation ${\cal{PV}}_{bt} \approx \left  \langle \zeta \right \rangle$.
}
	\label{fig:PV}
\end{figure}

\subsection{Potential Vorticity}

From the full NH-QGE system the barotropic and baroclinic potential vorticity are given by 
\begin{equation}
\label{eq:bcbt}
{\cal{PV}}_{bt} =\left  \langle \zeta \right \rangle - \left\langle J \left [W, \frac{\Theta'}{\partial_{Z} \overline{\Theta}} \right ] \right\rangle,\quad
{\cal{PV}}_{bc}= \zeta' - J\left [W, \frac{\Theta'}{\partial_{Z} \overline{\Theta}} \right ]'  +\partial_{Z}\left ( \frac{\Theta'}{\partial_{Z} \overline{\Theta}}\right ).
\end{equation}
Details of the characteristics of these two quantities are illustrated in Fig.~\ref{fig:PV}, with results for aspect ratios 1:2 and 1:5.  A snapshot of ${\cal{PV}}_{bt}$ as a function of $y$  indicates it is dominated by the barotropic vorticity $\left  \langle \zeta \right \rangle$. 
This signal oscillates in $y$ reflecting the spontaneous generation of a Kolmogorov-like flow. Consistent with the inverse energy cascade we find the magnitude of this signal is unbounded in time. In contrast, the baroclinic contribution to ${\cal{PV}}_{bt}$ remains bounded and fluctuates randomly about zero without coherence. This result suggests that the barotropic dynamics is essentially linear. In contrast ${\cal{PV}}_{bc}$ and its component terms, which are of roughly equal magnitude, vary rapidly about zero mean and saturate in time (not shown). 
The finding that dissipation and forcing are both weak suggests that the barotropic manifold is amenable to the application of equilibrium statistical mechanics \citep{bouchet2012}. 

\begin{figure}
\centering
\begin{tabular}{ll}
\subfloat{\includegraphics[width=0.4\textwidth]{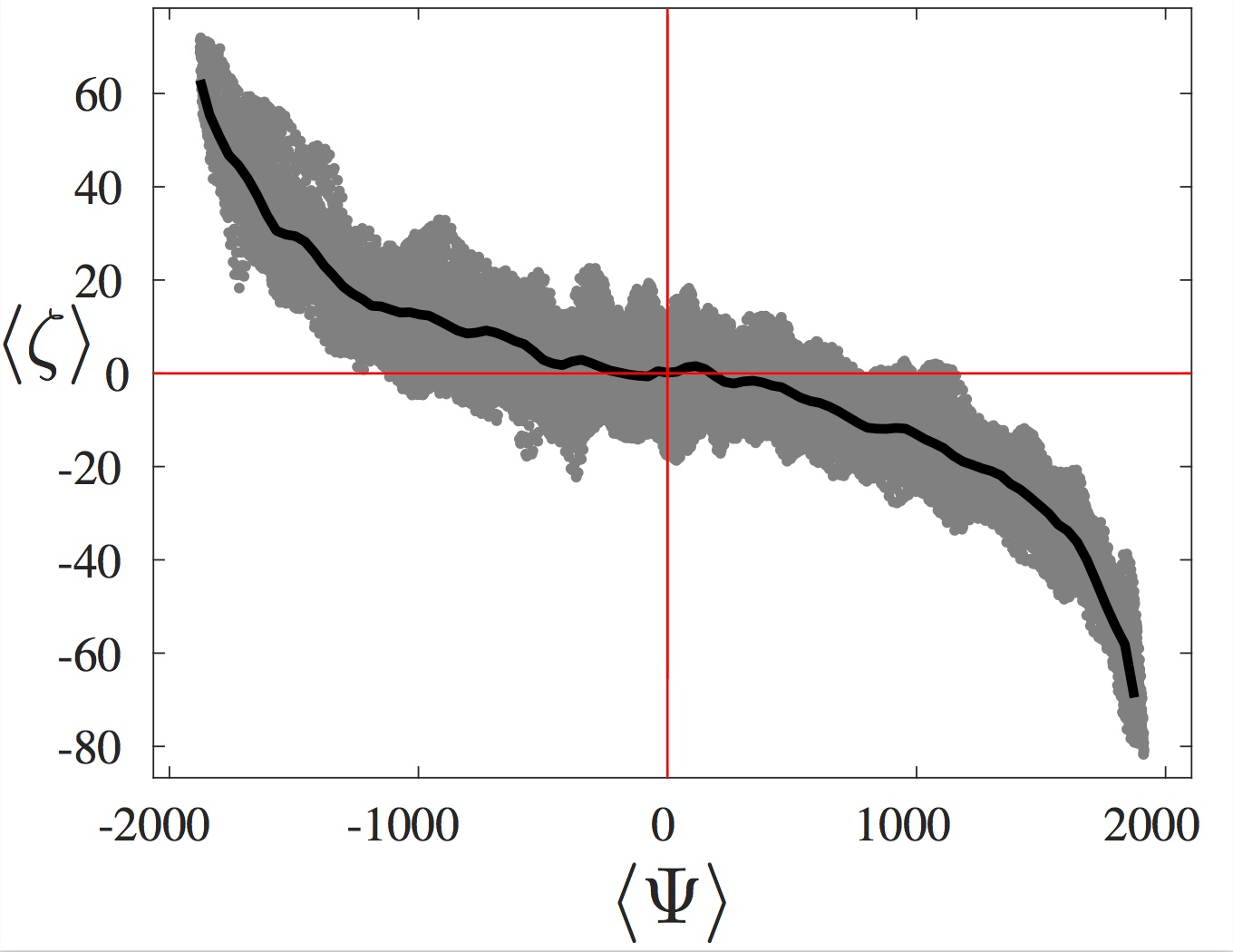}} & \hspace{1em}
\subfloat{\includegraphics[width=0.4\textwidth]{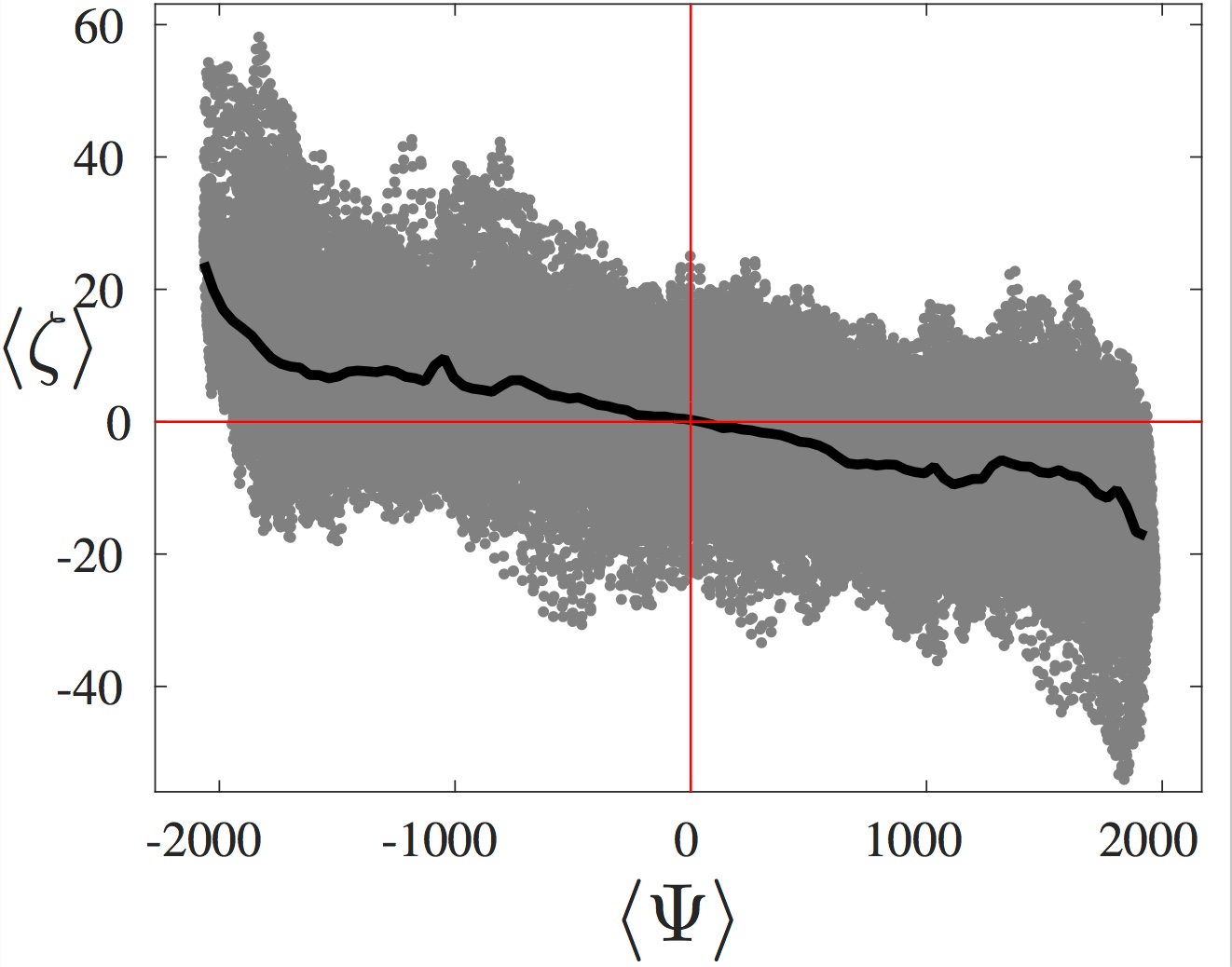}} 

\end{tabular}
	\caption{\small Scatter plots of $\left \langle \zeta \right\rangle$ vs $\left \langle \Psi \right\rangle$ for aspect ratios 1:1 (left) and 1:5 (right), suggesting a functional relation (black curve) obtained by pointwise averaging. The plots indicate $a_2a_4>0$. 
	}
	\label{fig:scat}
	\vspace{-1ex}
\end{figure}

\section{Conclusion}
\label{sec:Conc}

In this paper we have confirmed, following earlier work \citep{Julien12,Rubio14}, that geostrophic turbulence is unstable to the formation of large-scale vortices and investigated the properties of this state when the doubly periodic domain we use in the horizontal changes from square to rectangular. We have shown that with increasing domain anisotropy the large-scale vortex structure is replaced by a shear flow of Kolmogorov type parallel to the shorter side and superposed on the turbulent state. The flow has jet-like structure {with a well-defined characteristic scale comparable to the short box scale} that forms spontaneously. It is neither externally imposed nor the result of imposed external stirring \citep{Smith94,bouchet09,Laurie17} -- here the flow is maintained by 3D turbulent fluctuations that are determined self-consistently. The jets all have a well-defined mean separation that depends on the aspect ratio but undergo meander that may be intrinsic or driven by fluctuations. Evidently jet formation does not require the presence of a $\beta$ term.

These predictions resemble qualitatively the predictions from both numerical simulations of the damped 2D Navier-Stokes equations driven by imposed white noise \citep{Smith94,bouchet09,Laurie17} and those from equilibrium statistical mechanics for 2D flows \citep{bouchet2012}. Since the noise is imposed in the simulations it cannot respond to the box-scale structures that evolve. This is also the case in the equilibrium statistical description where noise must be assumed to be present to drive the system to equilibrium but the equilibrium reached is independent of the noise. This approach is based on maximizing entropy, defined in terms of the vertical vorticity, subject to constraints derived from the inviscid equations of motion. The variational problem leads to a monotonically increasing relation between the most probable values of the vorticity ${\bar\zeta}$ and streamfunction ${\bar\Psi}$. We assume that these values are those realized by the flow, and hence drop the overbars in the discussion that follows. Whether the predictions of this type theory are relevant to a 3D forced dissipative flow such as geostrophic turbulence remains a question, however. Figure~\ref{fig:scat} provides evidence that a $\zeta(\Psi)$ relation in fact exists for both $\Gamma=1$ and $\Gamma\ne1$ and is qualitatively similar to that found in 2D hydrodynamics. Indeed if we suppose, following \citep{bouchet09,bouchet2012}, a relation of the form $\zeta=a_2\Psi+a_4\Psi^3+...$ we conclude from Fig.~\ref{fig:scat}(a) that in both cases $a_2<0$, $a_4<0$. The nonzero value of $a_4$ is important in determining the nature of the transition from the box-scale dipole present when $\Gamma=1$ to the jet-like flow parallel to the shorter side present when $\Gamma\gtrsim1$, a transition whose presence here accords with both simulations of the damped 2D Navier-Stokes equations driven by noise \citep{bouchet09,Laurie17} and the prediction from equilibrium statistical mechanics \citep{bouchet09,bouchet2012}. The latter predicts the presence of a phase transition from the dipole state to a parallel shear flow already at small values of the elongation $\Gamma-1$ \citep{bouchet09,bouchet2012}), a result qualitatively similar to both the 2D simulations and our self-consistent 3D simulations in Fig.~\ref{fig:allapsectratios} (second row: $\Gamma=1.1$).

While the correspondence between our numerical results and equilibrium statistical mechanics theory is encouraging it is also evident that our system is not an equilibrium system -- it is a forced dissipative system. In this case, as already discussed, we expect to see an inverse energy cascade towards large scales and a direct entrophy cascade to small scales. However, if both the forcing and dissipation are appropriately weak on the large scales of interest, which we believe to be the case, the resulting nonequilibrium states that are observed are nevertheless expected to be close to the equilibrium states identified in the equilibrium statistical mechanics approach, a prediction corroborated in direct numerical simulations of a stochastically driven vorticity equation in 2D \citep{Smith94,bouchet09,Laurie17}. In particular, the transition from a dipole flow to a parallel shear flow with increasing elongation, predicted by the statistical mechanics approach, persists into the nonequilibrium regime (see Fig.~23 of \citet{bouchet2012}).

The above theory \citep{bouchet2012} has been developed for 2D flows. Our system is in contrast fully 3D, at least on small scales. However, the large-scale barotropic mode studied here obeys 2D dynamics driven by stochastic baroclinic forcing, and therefore shares many of the properties of these flows discussed by \citet{bouchet2012} and \citet{Laurie17}. In particular we expect that our flows should also undergo a transition with increasing anisotropy from a dipole flow to a parallel shear flow, a prediction confirmed in our simulations. We emphasize that these are performed on an asymptotically reduced model valid in the limit ${\rm Ro}\to 0$ \citep{Julien12,Rubio14}. Given that the predictions of this model have been confirmed in subsequent simulations of the primitive equations at $E=10^{-7}$ \citep{Stellmach}, $E=10^{-6}$ \citep{bF14} and $E=5\times 10^{-6}$ \citep{cG14} we expect that our findings also apply to NSE at sufficiently low Ekman numbers. That this is in fact the case is demonstrated by \cite{cG17} who show that when $E=10^{-5}$ jets also appear when the elongation exceeds 10\%. {The characteristic horizontal scales of the multi-jet states for larger elongations are also comparable.} However, there are some differences, too, in that Guervilly and Hughes also observe quite long-lived cyclonic vortices within their jets, whereas our system exhibits both cyclonic and anticyclonic vortices in equal numbers. This is a consequence of the reflection symmetries $x\to -x$, $y\to -y$ of the NH-QGE system that are present at leading order in the limit ${\rm Ro}\to 0$ ($E\to 0$).

The above discussion indicates that the formation of large-scale structures, be they box-scale vortices or jets, is a very robust phenomenon, independent of the details of the fluctuations driving the system, independent of the specific system studied and even independent of its dimensionality, provided only that that the flow is strongly anisotropic. Thus large-scale vortices are also present in 3D nonrotating systems, provided the fluid layer is sufficiently thin, thereby forcing the turbulent flow to be anisotropic on large scales \citep{xia2008,xia2009}. We expect that in the presence of doubly periodic boundary conditions this system will also undergo a transition with increasing domain anisotropy from the large-scale vortex state observed in square domains to a parallel shear flow, although this is of course difficult to confirm in experiments carried out in bounded domains. In thicker layers the large-scale vortex reduces vertical motion rendering the system susceptible to upscale energy cascade \citep{xia2011} and reinforcing the vortex. In this case, too, we expect a transition to large-scale shear flow with increased anisotropy much as occurs in convectively driven turbulence \citep{goluskin2014}. Both conjectures are amenable to confirmation by direct numerical simulation.

{\bf Acknowledgment}
This work was supported in part by the National Science Foundation under grants DMS-1317666, EAR-1620649 (KJ), DMS-1317596 (EK) and the NASA Earth and Space Science Fellowship Program (MP). The authors are grateful to C. Guervilly and D.W. Hughes for sharing their results prior to publication.

\bibliographystyle{jfm}
\bibliography{mybib}

\end{document}